\documentclass[doublecol]{epl2} 
\usepackage{amsmath}
\usepackage{amssymb}
\usepackage{amsthm} 
\usepackage{bm}
\usepackage{color}

\newcommand{\ket}[1]{\left|#1\right>}
\newcommand{\bra}[1]{\left<#1\right|}
\newcommand{\nn}{\nonumber\\}

\newcommand{\bea}{\begin{eqnarray}}
\newcommand{\ea}{\end{eqnarray}}
\newcommand{\eea}{\end{eqnarray}}

\newcommand{\sumint}[1]

\title{Identifying strongly correlated 
supersolid states on the optical lattice by quench-induced 
$\pi$-states}

\author{Uwe R. Fischer and Bo Xiong}
\shortauthor{}

\institute{                    
  Seoul National University,   Department of Physics and Astronomy, Center for Theoretical Physics \\
151-747 Seoul, Korea}


\abstract{We consider the rapid quench of a one-dimensional strongly correlated supersolid 
to a localized density wave (checkerboard) phase, and calculate the first-order coherence signal following the quench. It is shown that 
unique coherence oscillations between the even and odd sublattice sites of the checkerboard
are created by the quench, which are absent when the initial state is described by a Gutzwiller product state. 
This is a striking manifestation of the versatility of the far-from-equilbrium and nonperturbative collapse and revival phenomenon
as a microscope for quantum correlations in complex many-body states. For the present example, this 
opens up the possibility to discriminate experimentally between mean-field and many-body origins of supersolidity.}

\begin{document}

\maketitle

The single-particle density matrix, from which the first-order coherence 
of the system is defined, represents 
a sensitive probe of the correlation properties of interacting quantum many-body systems.
This becomes especially apparent in far-out-of-equilibrium dynamics when one rapidly (nonadiabatically) quenches from one quantum phase to another, 
which has recently led to intensive research into the properties of the resulting
{\em dynamical} quantum phase transitions \cite{OutofEq}. 
By now the most familiar nonequilibrium phenomenon occurring in such transitions 
are the collapse and revival oscillations of first-order coherence, created after sweeping from the superfluid 
to the Mott phase in optical lattices. When only conventional contact interactions are present, 
this has been first observed in  the classic paper \cite{Greiner}.
Experiments on the superfluid--Mott transition are now entering a high-precision age \cite{Bakr}, enabling the detection of many-body correlation functions and their temporal behavior on the single-particle and single-site accuracy level.  
Stimulated {\it inter alia} by this fact, various facets of the collapse and revival phenomenon were theoretically investigated recently in a number of papers, e.g. in \cite{CRTheory}. 

The extended Bose-Hubbard model, realizable
experimentally via cold atoms or molecules with large dipole moments 
stored in optical lattices \cite{Trefzger}, 
leads to a significant extension of the phase diagram of bosons on an optical lattice, 
including supersolid phases. By definition, for supersolids, the order parameter 
amplitude, which is homogeneous for a superfluid ground state, 
becomes spatially modulated, caused in the Bose-Hubbard model by 
two-body interactions coupling neighboring lattice sites with each other. 
While the implementation of supersolids in optical lattices cannot answer the intensely
debated question whether they occur in {\em real} solid-state systems like $^4$He, where  
the existence of  (growth-induced) defects in the crystal complicates the interpretation of experiments \cite{Andreev,Thouless,LeggettSuperSolid,SSNature,SSDoubts}, the microscopic and thus many-body origin 
of the supersolid phase can be understood better in such an artifically engineered system.  

In the following, we derive a unique signature for the imprint of the 
initial many-body correlations contained in the supersolid state from the far-out-of-equilibrium collapse and revival oscillations of the extended Bose-Hubbard model after a quench. The quench envisaged is the analog of the superfluid--Mott quench 
in that a delocalized phase-coherent (here supersolid) phase is quenched to an incoherent and localized, here density wave (DW) phase. 
We shall demonstrate that the final time-dependent coherence signal can be used as a sensitive probe of the initial strongly correlated 
supersolid state and 
therefore of a possible many-body origin of supersolidity. 
We show that even for very small density modulations of the initial 
supersolid ground state, which might be difficult to detect directly in experiment 
\cite{Scarola,Ye}, a distinct first-order coherence signal appears after the quench, 
providing a unique signature of supersolidity caused by both nonlocal interaction 
couplings and strong correlations.

We consider here that quantum fluctuations are strong in the initial ground state, 
for a quasi-one-dimensional (quasi-1D) Bose gas in a 1D optical lattice directed along the axis \cite{Monien,Mathey}. We determine  the ground state by exact diagonalization (ED), and  contrast the correlation response after the quench thus obtained  with the temporal correlations when the initial ground state is assumed to be describable by a Gutzwiller product state (GW) \cite{Trefzger}. 
It is stressed that we propagate both ED and GW initial wavefunctions exactly.
{We note in this regard that applying the {\em time-dependent} GW mean-field approach, used 
to describe small oscillations on top of the ground state, cf.\,\cite{Kovrizhin},   
to situations far from the (instantaneous) ground state (e.g. for collapse and revival oscillations), leads 
to both quantitatively and qualitatively incorrect predictions in the extended 
Bose-Hubbard model \cite{FX}.
By exact propagation, we conclusively demonstrate the strong dependence of the dynamical correlation response after the quench on the presence of many-body correlations in the initial state.} The final 
time-dependent correlation functions can therefore serve as a sensitive probe of whether the initial state correlations are correctly captured by a specific ground-state wavefunction
ansatz for the supersolid state.

 
The extended Bose-Hubbard Hamiltonian, describing bosons with dipolar interactions in the lowest Bloch band of an optical lattice, contains the nearest neighbor tunneling rate $J$, onsite interaction coupling $U$, and finally the nearest neighbor (NN) coupling $V$. Assuming 
a homogeneous system (no external trapping potential), it reads
\bea
\hat H &=& - J  \sum_{<ij>}^M \hat a_i^\dagger \hat a_j  
+ \frac U2 \sum_i^M (\hat n_i-1) \hat n_i 
+ \frac V2 \sum_{<ij>}^M {\hat n_i \hat n_j} ,
\nn
\label{H}
\ea 
where $M$ denotes the number of sites and $<\!ij\!>$ indicates a summation over NN 
(we will provide below a discussion of the influence of including off-site interaction couplings beyond NN). 
{The onsite and offsite couplings $U$ and $V$ are proportional to the dipole moments
squared and depend on integrals over products of the Wannier orbitals for the lowest Bloch band. A detailed derivation of the extended Bose-Hubbard Hamiltonian for dipoles in an optical lattice can be found, for example, in \cite{Trefzger}.} 
We assume the dipoles to be oriented perpendicular to the axial direction such 
that the NN coupling is repulsive, $V>0$. We use $U\equiv 1$ as unit of the energy and $2\pi/U$ as the unit of time in what follows ($\hbar\equiv 1$); the off-site coupling is fixed at $V=0.9$; 
this choice of $V\lesssim U$ is motivated by the fact that then  
small population imbalance supersolids can be obtained at 
relative tunneling rates $J/({\bar nU})\ll 1$, also cf.\,\cite{Iskin}.  

Previous studies into the extended Bose-Hubbard model, covering the whole range of fillings 
and couplings, have found various other phases like superfluids and localized Mott and DW phases   
besides supersolids \cite{Batrouni,Mishra}.  
We concentrate in what follows on supersolid initial states and final (after the quench) 
DW states.

The ground-state density modulation in the supersolid state for the lattice model \eqref{H} is related to a 
structural instability, developing when the excitation spectrum above the superfluid ground state, at the transition to the supersolid state, touches the quasi-momentum axis at $q=\pi/a$ \cite{Kovrizhin,Schuetzhold,Buehler}. 
This leads to 
``A'' (with indices $i=A=1,3,5,\ldots$) and ``B''  (with indices $i=B=2,4,6,\ldots$)
sublattices following a checkerboard pattern associated with the density modulation of the supersolid ground state at wavevector $q=\pi/a$.  
A single scalar measure distinguishing the supersolid state from the superfluid state 
is therefore the population imbalance 
\bea 
{\mathcal I}_{\rm SS}= \frac{|\bar n_A- \bar n_B|}{\bar n_A+ \bar n_B}, 
\label{I_SS}
\ea
where $\bar n_A$ and $\bar n_B$ are average fillings of sublattice sites; 
hence ${\mathcal I}_{\rm SS}=0$ for a superfluid and ${\mathcal I}_{\rm SS}\neq 0$
for a supersolid state.

\begin{center}
\begin{figure}[t]
\centering
\includegraphics[width=.48\textwidth]{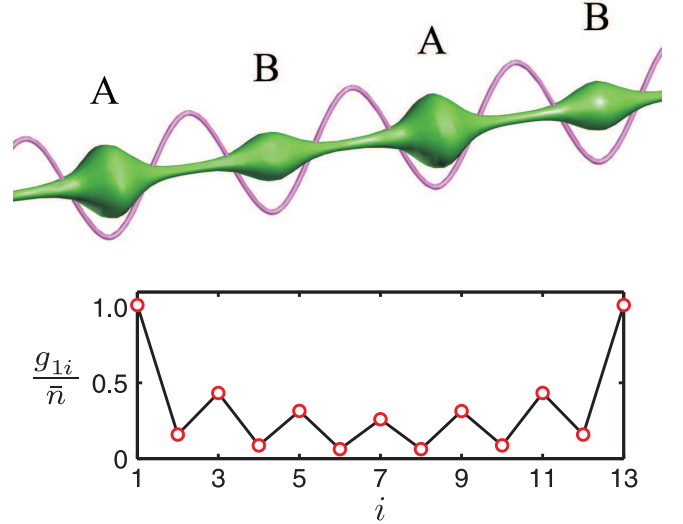}
\caption{\label{initialgij} 
Top: Artist's view of the supersolid ground state in a laterally tightly confined quasi-1D gas with axially 
superimposed optical lattice (solid line). For representation purposes, the density imbalance 
${\mathcal I}_{\rm SS}$ is chosen large.
Bottom: Correlation function $g_{1i}/\bar n$ of initial state for $M=12$ 
and $N=8$ 
from ED with $\bar n=2/3$ and $J=0.05$ 
(open circles); solid lines are a guide to the eye. }
\end{figure}
\end{center} 

The phase and modulus of the single-particle density matrix 
\bea 
\sigma_{kl} = \langle \hat a_k^\dagger \hat a_l \rangle 
= |\langle \hat a_k^\dagger \hat a_l \rangle | \exp[i\theta_{kl}]
\label{defsigma}
\ea
embody density and phase correlations between sites, and 
lead to first-order correlations 
$g_{ij}= \Re [\sigma_{ij}] = 
|\langle \hat a_i^\dagger \hat a_j \rangle | \cos[\theta_{ij}]$.
We calculated the ground-state correlation function corresponding to the Hamiltonian \eqref{H}
for a supersolid state in a lattice with 
$M=12$ sites and $N=8$ respectively $N=16$ particles, i.e., at 
fillings $\bar n =2/3$ and $\bar n =4/3$. 
For this system size, the determination of both initial ground state and temporal evolution can be performed by ED routines, employing periodic boundary conditions.   
{In addition, we have verified the qualitative robustness of our results  by comparing with sample calculations for $M=10$ sites at fillings $\bar n =4/5$ as well as  $M=14$ sites at filling $\bar n =9/14$ 
for ground state and quench dynamics, and $M=16$ sites at $\bar n =5/8$ for the supersolid ground state.

There is the inherent lattice symmetry leading to degenerate SS states which have either $n_A>n_B$ or $n_A<n_B$.
We choose one of these states, say with $n_A>n_B$, by the following procedure. We perform a large sequence of 
numerical calculations with different initial seeds for many-body states in the ED calculation, to obtain a statistical distribution of various values of ${\cal I}_{\rm SS}$ with different degenerate states. From this distribution, we choose the initial 
state as the one with the most probable value of ${\cal I}_{\rm SS}$ which fulfills $n_A>n_B$. }

As a result, we obtain for the supersolid ground state that, at constant phase $\theta_{jk}$, 
first-order correlations are strictly positive,
decaying with distance in the power-law fashion characteristic of low-density quasi-1D Bose gases 
\cite{Efetov}, cf.\,Fig.\,\ref{initialgij}.

\begin{center}
\begin{figure}[tb]
\centering
\includegraphics[width=.485\textwidth]{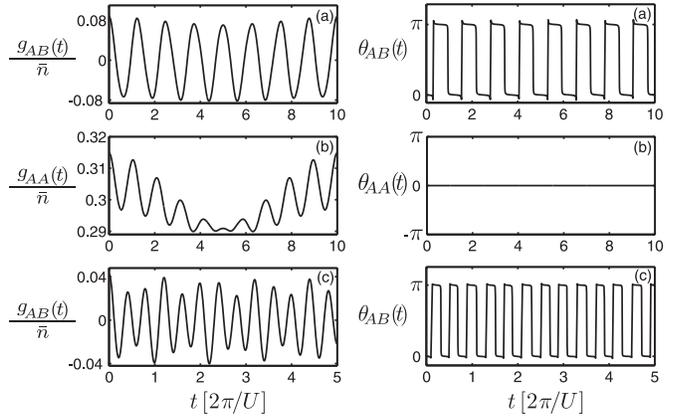}
\caption{\label{pi-state}
Coherence and phase-difference signals obtained in ED (we fix $i=1,j=4$ in $g_{AB}$
and $i=1,j=5$ in $g_{AA}$). 
Left column: $g_{AB}/\bar n$ (a) and $g_{AA}/\bar n$ (b) for filling $\bar n=2/3$  at $J=0.05$ (i.e. with initial correlations shown in Fig.\,\ref{initialgij}). Right column: Corresponding phase difference signals.  
(c) shows the $g_{AB}$ and $\theta_{AB}$ results for higher filling $\bar n=4/3$ and at $J=0.08$ 
(note the different time axis). Imbalance $\mathcal I_{\rm SS}=0.0136$ in (a),(b) 
and $\mathcal I_{\rm SS}=0.009$ in (c).}
\end{figure}
\end{center}

Next we let the tunneling abruptly go to zero in a quasi-instantaneous quench, $J\rightarrow 0$, which would, in an adiabatic transition, transfer the supersolid to the localized and 
insulating DW phase.
The main feature we observe is that there are 
two different coherence signals obtained after the quench, depending on the distance between sites 
considered being $|i-j|=$ even or odd. 
This is a first crucial difference from the uniform (long-range) collapse and revival in the superfluid-Mott quench \cite{Greiner,CRTheory}.
The ``AA'' (and equivalently ``BB'') correlations between the sites belonging to the same sublattice ($|i-j|=$\,even) show (partial) collapse and revival oscillations, i.e. with strongly reduced amplitude, 
at constant phase $\theta_{AA}$ ($\theta_{BB}$), cf. Fig.\,\ref{pi-state}(b).
On the other hand, the ``AB'' correlations belonging to sites of different sublattices ($|i-j|=$\,odd) show periodic changes from positive to negative values of correlations, see Fig.\,\ref{pi-state}, while the average populations $\bar n_A$ are $\bar n_B$ are both constant 
in time (implying that the phase difference switching does not lead to a redistribution of occupation numbers). 
The phase difference $\theta_{AB}$ in the latter case thus switches from values close to zero to close to $\pi$, for prolonged time periods of order the ordinary revival time scale $2\pi/U$
[we are using a representation of the phase difference which takes into account that 
it is defined modulo $2\pi$, i.e. in particular identify $2\pi$ and 0].
The first order correlation function $g_{AB}(t)$  
goes through zero when $\theta_{\rm AB}$ switches from zero to $\pi$
and vice versa. For these out-of-equilibrium states far from the DW ground state, we coin 
the term $\pi$-{\em states} \cite{Raghavan}.


We stress that for an ordinary superfluid to Mott quench at $V=0$, no $\pi$-state is obtained, 
even in the present strongly correlated low-filling 1D lattice case. Hence a 
nonvanishing off-site coupling $V$ in the Bose-Hubbard
Hamiltonian \eqref{H} is crucial for obtaining the large negative AB correlations characteristic of the
$\pi$-state. 
Furthermore, we also observed that when the initial imbalance ${\mathcal I}_{\rm SS}$ 
becomes large (achieved when we take $V>U$ at small $J$), and therefore the initial supersolid state is less strongly correlated across different sites, 
that the tendency to form a $\pi$-state disappears. 
In the extreme case of ${\mathcal I}_{\rm SS}\rightarrow 1$ (one sublattice 
empty), the phase simply changes linearly, $\theta_{AB}\propto t$.  

\vspace*{-2em}

We now contrast the occurrence of $\pi$-states, obtained starting from a strongly correlated initial state, with the first-order correlation signal from an initial supersolid state assumed to be given in GW mean-field theory. The many-body state can then be written as the product state \cite{Iskin}
\begin{equation} \label{GWSSinit}
           |{\rm GW} \rangle =  
           \prod_{A,B}\sum_{n_{A},n_{B}} 
           f_A (n_{A}) f_B (n_{B}) |n_{A}, n_{B}\rangle,
     \end{equation}
     where $\prod_{A,B}$ is a shorthand for taking the product over all sites, 
and the indices $A$ and $B$ run over all $M/2$ corresponding sites of either sublattice. 
The sublattices are separated here explicitly, with $f_A$ and $f_B$ in general 
being different distribution functions for either sublattice.

\begin{widetext}
\begin{multline}     
       \langle \hat{a}_{A}^{\dag} \hat{a}_{B} \rangle   
           =  \sum_{n_{A}, n_{B}} f_A (n_{A}) f_B (n_{B})
                   f_A^{*}(n_{A}+1) f_B^{*} (n_{B}-1) 
                   \sqrt{n_{B} (n_{A} + 1)}e^{itU(n_{A} - n_{B})}   
        \\ 
        \times  
        \prod_{\rm{NN}\,\it{l}}\sum_{n_l}|f_{(A,l)}|^{2} |f_{(B,l)}|^{2}
          e^{itV(\sum_{[Al]}n_{l} - \sum_{[Bl]}n_{l})} .
       \label{gABGW}
\end{multline} 
\end{widetext}

When the initial state is given in the product form \eqref{GWSSinit}, the {\em exact} time evolution of AB correlations in the limit of zero tunneling, $J=0$, after the quench can be written in the form of Eq.\,\eqref{gABGW} (with a similar expression for the AA and BB correlations), where NN\,$l$ designates all sites 
$l$ which are nearest neighbors to the sites with given indices $A$ and $B$, and we assume for simplicity that the distance $|A-B|> 2$ \cite{FX}. 
The notation $[Al]$ and $[Bl]$ indicates sums over all NNs $l$ at the sites $A$ and $B$, respectively, and $f_{(A,l)}$ means $f_A(n_l)$ where $l$ is a NN to $A$ and for $f_{(B,l)}$ correspondingly.   
We therefore see that in case the initial state can be written in Gutzwiller product form \eqref{GWSSinit}, the final-state time-dependent correlations solely depend 
on the on-site initial number distribution functions in the A,B sites, and in 
particular on their widths and centers.

\begin{center}
\begin{figure}[hbt]
\centering
\includegraphics[width=.49\textwidth]{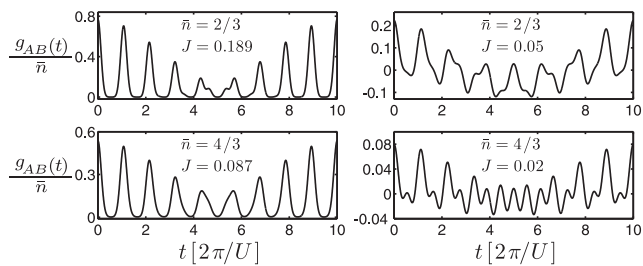}
\caption{\label{GWSSprop} Coherence signal $g_{AB}(t)$
between sublattices obtained when propagating exactly a GW supersolid initial state
after a quench to $J=0$, 
for the indicated values of filling and tunneling [cf. for the 
ED results in Fig.\,\ref{pi-state} (a) and (c), left column].    
Top left has the same $\mathcal I_{\rm SS}=0.0136$ 
like the ED result of Fig.\,\ref{pi-state}(a), while on the right
$\mathcal I_{\rm SS}=0.5339$. 
Bottom for larger filling, left again $\mathcal I_{\rm SS}=0.0136$
and  right ${\mathcal I}_{\rm SS}= 0.25$.}
\end{figure}
\end{center} 

We time evolve the initial GW state \eqref{GWSSinit}, found upon minimizing 
$\bra{\rm GW} \hat H - \mu \hat N \ket{\rm GW}$, where $\mu$ is the chemical potential, 
using the exact formula \eqref{gABGW}, and 
show the results in Fig.\,\ref{GWSSprop}. 
While there is some negative amplitude of $g_{AB}$ in the 
large-imbalance low-filling case (cf. the right upper plot in Fig.\,\ref{GWSSprop}), 
the highly symmetrical $\pi$-state of Fig.\,\ref{pi-state} (a) and (c) is not obtained.
For illustration purposes, we display in Fig.\,\ref{GWSSprop} the evolution 
results for the same low fillings like in Fig.\,\ref{pi-state}, for which the GW approach naturally cannot be expected to give the correct ground state, in particular for small population imbalance.
Generally speaking, a much larger $J$ is necessary to obtain in GW 
the same ${\mathcal I}_{\rm SS}$ as compared to ED at the same filling, 
which emphasizes the strong correlation of a low-filling supersolid in 1D. 
We have furthermore confirmed that the GW correlation response is predominantly  
determined by the imbalance ${\mathcal I}_{\rm SS}$, and at the 
same values of the latter quantity remains qualitatively  
similar at larger filling $\bar n$. We also note that the maximally
possible sublattice imbalance for the supersolid ground state decreases with 
increasing $\bar n$, so that at larger filling, where the non-number-conserving
GW theory becomes asymptotically exact, the tendency of developing negative
$g_{\rm AB}$ as shown in the right row of Fig.\,\ref{GWSSprop}
vanishes completely.
%
The fact that with an initial 
GW product state large negative AB correlations are not obtained 
serves to illustrate that both large nonlocal interaction coupling and strong correlations are necessary to obtain the large negative values of $g_{AB}$ shown in Fig.\,\ref{pi-state}
and thus an (almost) perfect $\pi$-state.

\begin{center}
\begin{figure}[hbt]
\centering
\includegraphics[width=.45\textwidth]{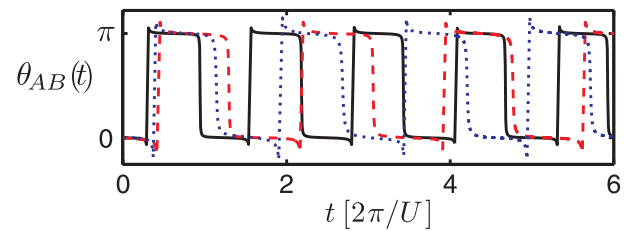}
\caption{\label{cutofflength} $\pi$-state switching of the relative sublattice phase $\theta_{AB}$ 
for different cutoff lengths $\Lambda$ of a dipolar off-site interaction. $J=0.05$, $\bar n =2/3$ ($M=12,N=8$); 
Solid black line $\Lambda=1$, dashed red line $\Lambda=2$, dotted blue line $\Lambda =3$.} 
\end{figure}
\end{center}

We also investigated in addition whether the $\pi$-state 
is robust under increasing the number of 
off-site interaction couplings taken in the extended Bose-Hubbard Hamiltonian \eqref{H}.  
For a dipolar interaction, the off-site interaction Hamiltonian for polarized dipoles
reads 
\bea 
{\hat H}_{\rm dd} = \frac V2 \sum_{ij}^M \frac{\hat n_i \hat n_j}{|i-j|^3}. 
\ea
We are truncating the coupling summation at the distance $|i-j|=\Lambda$, where
the NN off-site coupling in \eqref{H} corresponds to $\Lambda = 1$.
The ED results for the time dependent $\theta_{AB}$, 
for three different cutoff lengths, and the identical parameter set of Fig.\,\ref{pi-state} (a), are shown in Fig.\,\ref{cutofflength}. 
We see that the $\pi$-state, consisting in the switching of the relative AB phase between 0 and $\pi$, still occurs, although with a different period.  
{The period first increases when going from $\Lambda=1$ to $\Lambda=2$, and then decreases again for $\Lambda=3$. This behavior is due to the ABAB sublattice structure of the supersolid initial state.
The $\pi$-state period thus encodes nontrivial information on the 
interaction range chosen for the extended Bose-Hubbard model in a quench 
with a periodically modulated initial state.}

\begin{center}
\begin{figure}[hbt]
\centering
\includegraphics[width=.4\textwidth]{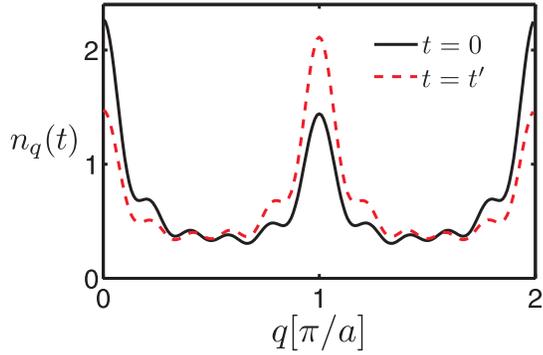}
\caption{\label{nq} {Quasi-momentum distribution function $n_q(t)$ for two different times.
The time $t'$ is defined by the influence of the $\pi$-state 
on the quasi-momentum distribution being most significant, corresponding to the negative minima of the correlation function $g_{AB}$. Parameters
are chosen identical to those of Fig.\,\ref{pi-state}\,(a).}}
\end{figure}
\end{center}

{The $\pi$-states can be detected experimentally, for example, by the time-dependence of peaks in the 
quasi-momentum distribution function 
\bea
n_q(t) =\frac1M \sum_{k,l} \langle \hat a_k^\dagger \hat a_l \rangle  e^{iqa(k-l)},
\ea 
which correspond to the visibility of interference peaks 
in the real-space density 
detected by time-of-flight experiments \cite{Scarola}. 
Pronounced peaks are located at $q=0$ and $q=\pi/a$, cf.\,Fig.\,\ref{nq}, 
whose time dependence 
should be easily discernible.
This is due to the fact that $g_{AA}(t)$ [$g_{BB}(t)$] and $g_{AB}(t)$ strongly differ on average and have sufficient oscillation amplitude even for small population imbalance ${\mathcal I}_{SS}$, cf.\,\,Fig.\,\ref{pi-state}, which leads to significant temporal variation and amplitude of $n_q(t)$. As shown in Fig.\,\ref{nq}, the appearance of the $\pi$-state leads to a 
sizable increase of the peak at $q=\pi/a$, while it decreases the $q=0$ peak.} 

We have presented a study of 1D supersolids on optical lattices undergoing a quench
to a density wave phase, and observed that a final nonequilibrium $\pi$-state 
signature develops, characteristic of strong correlations in the initial state. 
By contrast, the $\pi$-state does not occur when the inital state is described 
by a Gutzwiller product state ansatz.
This is a prime example of the high sensitivity of far-out-of-equilibrium 
many-body dynamics on initial correlations.
{As a corollary, even when a Gutzwiller product state can reasonably well emulate certain properties 
of the equilibrium ground state, 
it can potentially fail tremendously as an initial state for many-body evolution.} 
Correlation functions after a quench can therefore serve as a ``microscope'' 
of the initial correlations, in which the far-out-of-equilibrium conditions magnify the many-body character
of a state or the lack thereof.
In the present case, this offers the possibility to detect 
a strongly correlated supersolid state of matter even at very small density modulations, and thus to determine accurately the phase transition point to the 
superfluid state in which no $\pi$-state occurs. 

\acknowledgments

This research was supported by the 
Brain Korea BK21 program, the NRF of Korea, grant 
Nos. 2010-0013103 and 2011-0029541, and the Seoul National University Foundation Research Expense.


\begin{thebibliography}{0}

\bibitem{OutofEq}  J. Dziarmaga, Adv. Phys. {\bf 59}, 1063 (2010); 
A. Polkovnikov, K. Sengupta, A. Silva, and M. Vengalattore, Rev. Mod. Phys. {\bf 83}, 863 (2011); M.\,P. Kennett and D. Dalidovich, Phys. Rev. A {\bf 84}, 033620 (2011). For a quench in a Luttinger liquid without optical lattice, see M.\,A. Cazalilla, 
Phys. Rev. Lett. {\bf 97}, 156403 (2006).

\bibitem{Greiner} M. Greiner, O. Mandel, T.\,W. H\"ansch,
and I. Bloch, 
Nature {\bf 419}, 51 (2002).

\bibitem{Bakr} W.\,S. Bakr {\it et al.}, 
Science {\bf 329}, 547 (2010);
J.\,F. Sherson {\it et al.}, 	
Nature {\bf 467}, 68 (2010);
C. Weitenberg {\it et al.}, 
Nature {\bf 471}, 319 (2011); 
M. Mark {\it et al.}, 
Phys. Rev. Lett. {\bf 107}, 175301 (2011).

\bibitem{CRTheory} U.\,R. Fischer and R. Sch\"utzhold, Phys. Rev. A {\bf 78},
061603(R) (2008); 
F.\,A. Wolf, I. Hen, and M. Rigol, Phys. Rev. A {\bf 82}, 043601 (2010); 
J. Schachenmayer, A.\,J. Daley, and P. Zoller, 	
Phys. Rev. A {\bf 83}, 043614 (2011); 
E. Tiesinga and P. R. Johnson, Phys. Rev. A {\bf 83}, 063609 (2011);  
M. Buchhold, 
U. Bissbort, S. Will, and W. Hofstetter,
Phys. Rev. A {\bf 84}, 023631 (2011). 

\bibitem{Trefzger} {K. G\'oral, L. Santos, and M. Lewenstein, 
Phys. Rev. Lett. {\bf 88}, 170406 (2002);} C. Trefzger, C. Menotti, B. Capogrosso-Sansone,  and M. Lewenstein, 
J. Phys. B: At. Mol. Opt. Phys. {\bf 44}, 193001 (2011).

\bibitem{Andreev} A.\,F.~Andreev and I.\,M.~Lifshitz,
Sov.~Phys.~JETP {\bf 29}, 1107 (1969).

\bibitem{Thouless} D.\,J. Thouless,
Ann. Phys. (N.Y.) {\bf 52}, 403 (1969). 

\bibitem{LeggettSuperSolid} 
A.\,J.~Leggett,
Phys.~Rev.~Lett.~{\bf 25}, 1543 (1970).~

\bibitem{SSNature} E.~Kim and M.\,H.\,W.~Chan,
Nature {\bf 427}, 225 (2004); 
Science {\bf 305}, 1941 (2004). 

\bibitem{SSDoubts} 
A.\,J.~Leggett, 
Science {\bf 305}, 1921 (2004); for a recent alternative theoretical explanation of the $^4\!$He experiments see A.\,F. Andreev,  
J. Low. Temp. Phys. {\bf 168}, 126 (2012).


\bibitem{Scarola}   
V.\,W.~Scarola, E.~Demler, and S.~Das Sarma, 
Phys.~Rev.~A {\bf 73}, 051601(R) (2006).

\bibitem{Ye} J. Ye {\it et al.}, 
Phys. Rev. A {\bf 83}, 051604(R) (2011). 
  

\bibitem{Monien}   T.\,D. K\"uhner, S.\,R. White, and H. Monien, 
Phys. Rev. B {\bf 61}, 12474 (2000). 

\bibitem{Mathey} We remark that strongly correlated 1D supersolids   
have been discussed for {contact} interactions as well, 
e.g. by L. Mathey, I. Danshita, and C.\,W. Clark, Phys. Rev. A {\bf 79}, 011602(R) (2009) and A. Lazarides, O. Tieleman, and C. Morais Smith, Phys. Rev. A {\bf 84}, 023620 (2011).



\bibitem{Kovrizhin} 
D.\,L. Kovrizhin, G. Venketeswara Pai, and S. Sinha, 
Europhys. Lett. {\bf 72}, 162 (2005).





\bibitem{FX} U.\,R. Fischer and B. Xiong,
Phys. Rev. A  {\bf 84}, 063635 (2011).

\bibitem{Schuetzhold} R. Sch\"utzhold, M. Uhlmann, and U.\,R. Fischer,
Phys. Rev. A {\bf 78}, 033604 (2008).

\bibitem{Buehler} A. B\"uhler and H.\,P. B\"uchler, 
Phys. Rev. A {\bf 84}, 023607 (2011). 


\bibitem{Iskin}  M. Iskin,   
Phys. Rev. A {\bf 83}, 051606(R) (2011).

{
\bibitem{Batrouni} G.\,G. Batrouni, F. H\'ebert, and R.\,T. Scalettar, 
Phys. Rev. Lett. {\bf 97}, 087209 (2006).
\bibitem{Mishra} T. Mishra {\it et al.},
Phys. Rev. A {\bf 80}, 043614 (2009).
}


\bibitem{Efetov} K.\,B. Efetov and A.\,I. Larkin, Sov. Phys. JETP {\bf 42}, (1975).

\bibitem{Raghavan} The $\pi$-states introduced here 
are not to be confused with the $\pi$-phase oscillations in 
nonlinear bosonic Josephson junctions, see   
S. Raghavan, A. Smerzi, S. Fantoni, and S.\,R. Shenoy,
Phys. Rev. A {\bf 59}, 620 (1999). 






\end{thebibliography}
\end{document}